# QUANTUM PHASE SPASE REPRESENTATION FOR DOUBLE WELL POTENTIAL

Dmytro Babyuk

*Department of Chemistry, University of Nevada, Reno, Reno Nevada 89557*

A behavior of quantum states (superposition of two lowest eigenstates, Gaussian wave packet) in phase space is studied for one and two dimensional double well potential. Two dimensional potential is constructed from double well potential coupled linearly and quadratically to harmonic potential. Quantum trajectories are compared with classical ones. Preferable tunneling path in phase space is found. An influence of energy of initial Gaussian wave packet and trajectory initial condition on tunneling process is analyzed.. Coupling effect on trajectories and preferable tunneling path for two dimensional potential is studied.

## I. INTRODUCTION

The quantum dynamical simulation of molecular processes remains difficult for many real systems despite advances in computational speed. Introduction of different approximations can reduce challenge. But any approximation has unavoidable drawback associated with impossibility to reproduce inherently quantum effects adequately. Before using an approximation one needs to know about correspondence between classical and quantum dynamics. One of the methods to learn is a comparison of quantum and classical behavior in phase space. But the uncertainty principle imposes limitations to definition of quantum mechanics in phase space. In other words, there is an ambiguity which gave rise to a variety of different quantum phase space distributions.

One of the widest used of them is introduction of Wigner transform [1,2]. It generates from coordinate-representation wave function the real valued one, named the Wigner function. It satisfies many classical-like conditions. But the Wigner function suffers the shortcoming that it is not everywhere positive and therefore cannot be a probability density in phase space. To overcome a drawback of Wigner method, an alternative approach using the Husimi function [3] has been developed. Although it provides a quantum phase space distribution which is everywhere positive, Husimi function does not satisfy all classical-like properties.

All previous efforts consist in seeking for an appropriate transformation from a coordinate representation function to a real valued phase space distribution. Qualitatively new approach proposed by Go.Torres-Vega and Frederick [4] is based on a formulation of quantum mechanics directly in phase space. According to their work, the quantum phase space representation (QPSR) postulates the existence of complete set of basis vectors |q,p>. Quantum state can be represented by wave function dependent on coordinate and momentum simultaneously. Operators of these variables are non-local thus obeying the uncertainty principle. They are constructed in such a way that the correspondence between the classical and quantum Liouville equation is satisfied the best. The Hamiltonian operator is then built from the momentum and coordinate operators, and thus the Schroedinger equation in phase space. The square magnitude of wave function can be compared with that of a classical density. By analogy to Hamiltonian description of classical mechanics, quantum trajectories can be defined from probability density and fluxes. These quantum trajectories are compared with classical ones. It is another way to analyze classical quantum correspondence.

Many problems have been studied by solving non-stationary Schroedinger equation in phase space for different potentials. If exact analytical solution for definite potential can be derived in coordinate representation, the same conclusion can be typically reached in phase space representation. Examples are - free particle [5], linear potential [6], harmonic potential [4,7] and Morse potential [8]. More complicated potentials for which Schroedinger equation does not have an analytical solution require diagonalization in basis set of known eigenstates for bound potential.

The purpose of the following work is to study the behavior of quantum state in phase space which does not have analytical solution, namely double well potential. This potential has a barrier, hence such phenomenon as tunneling can be explored in this sample. Some efforts have been made in [9,10]. We wish to represent this task in more details and in multi-dimensions.

The remainder of the paper is organized as follows. In Sec.II, we introduce general equation for the double well potential in the QPSR. Sections III and IV are concerned with application to one dimensional and two dimensional potential, respectively. Finally, we summarize our findings in Sec.V.

## II. QPSR FOR DOUBLE-WELL POTENTIAL

Quartic double well potential in coordinate representation is given by

$$U(q) = \alpha q^4 - \beta q^2 + E_0 \quad (2.1)$$

If one chooses constant $\beta$ as $\beta = 2\sqrt{\alpha E_0}$, where $E_0$ is a barrier height and $\alpha$ is any arbitrary constant, then the potential has a form, as shown in Fig.1.1a.

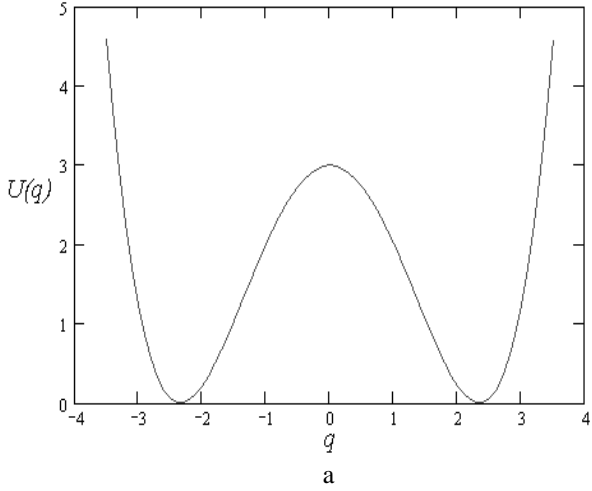

a

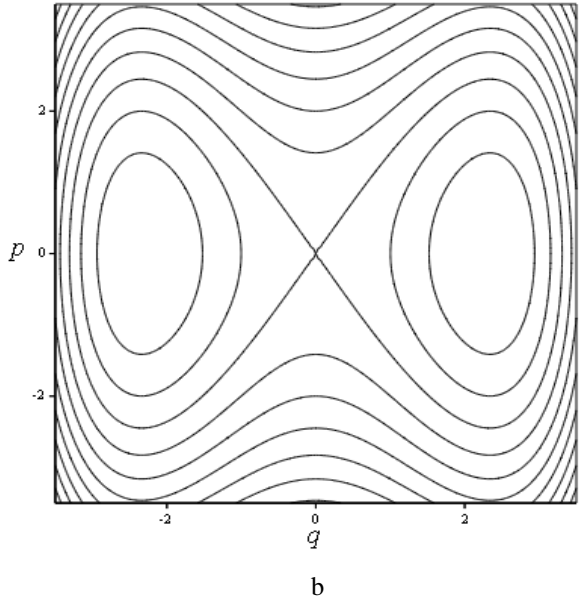

b

**Figure 1.1.** Double well potential as function of coordinate (a); contour plot of double well potential in phase space (b).

The distance between two wells in this case is $\sqrt[4]{\alpha/E_0}$. So $\alpha$ is a value which specifies distance between wells.

In the QPSR momentum and coordinate operators have the following form [4]

$$\hat{P} = \frac{p}{2} - i\frac{\partial}{\partial q}, \qquad \hat{Q} = \frac{q}{2} + i\frac{\partial}{\partial p} \quad (2.2)$$

Thus the Hamiltonian is constructed in the form

$$\hat{H} = \frac{1}{2}\hat{P}^2 + U(\hat{Q}) \quad (2.3)$$

Substituting (2.2) into (2.3) and taking into account (2.1) we have Hamiltonian for double well potential in phase space representation

$$\hat{H} = \frac{1}{2}\left(\frac{p}{2} - i\frac{\partial}{\partial q}\right)^2 + \alpha\left(\frac{q}{2} + i\frac{\partial}{\partial p}\right)^4 - 2\sqrt{\alpha E_0}\left(\frac{q}{2} + i\frac{\partial}{\partial p}\right)^2 + E_0 \quad (2.4)$$

Stationary Schroedinger equation $\hat{H}\Psi(p,q) = E\Psi(p,q)$ with the Hamiltonian (2.4) is solved by diagonalizaiton of Hamiltonian matrix in basis set of harmonic oscillator. For example, if potential parameters are $\alpha$=0.1 and $E_0$=3, then the six nearest eigenvalues are

$$E_0 = 0.992, \quad E_1 = 0.998, \quad E_2 = 2.595, \quad E_3 = 2.826,$$
$$E_4 = 3.828, \quad E_5 = 4.698$$

According to [4] general form of quantum Liouville equation in phase space is

$$\frac{\partial}{\partial t}<\Gamma|\rho|\Gamma> = -\frac{\partial}{\partial q}\frac{1}{2}\left[<\Gamma|\hat{P}\hat{\rho}|\Gamma> + <\Gamma|\rho\hat{P}|\Gamma>\right] + \frac{\partial}{\partial p}\sum_{n=1}^{\infty}V_n\sum_{l=0}^{n-1}<\Gamma|\hat{Q}^l\hat{\rho}\hat{Q}^{n-l-1}|\Gamma> \quad (2.5)$$

It follows from (2.5) that the quantum fluxes should have a form

$$Jq = \frac{1}{2}\left[<\Gamma|\hat{P}\hat{\rho}|\Gamma> + <\Gamma|\rho\hat{P}|\Gamma>\right]$$

$$Jp = -\sum_{n=1}^{\infty}V_n\sum_{l=0}^{n-1}<\Gamma|\hat{Q}^l\hat{\rho}\hat{Q}^{n-l-1}|\Gamma>$$

In case of double well potential they can be converted into

$$Jq = \text{Re}[<\Gamma|\hat{P}|\Psi><\Gamma|\Psi^*>] \quad (2.6)$$

$$Jp = -2\text{Re}\{-\beta<\Gamma|\Psi><\Gamma|\hat{Q}^*|\Psi^*> + \alpha[<\Gamma|\Psi><\Gamma|\hat{Q}^{*3}|\Psi^*> + <\Gamma|\hat{Q}|\Psi><\Gamma|\hat{Q}^{*2}|\Psi^*>]\} \quad (2.7)$$

In classical mechanics density fluxes are given by the product of probability density and velocity. Using this analogy to quantum mechanics one can formulate

$$Jp(p,q,t) = \rho(p,q,t)v_p = |\Psi(p,q,t)|^2 \frac{dp}{dt}$$
$$Jq(p,q,t) = \rho(p,q,t)v_q = |\Psi(p,q,t)|^2 \frac{dq}{dt} \quad (2.8)$$

System (2.8) can be rewritten in the form

$$\frac{dp}{dt} = F_1(p,q,t) = \frac{Jp(p,q,t)}{\rho(p,q,t)},$$
$$\frac{dq}{dt} = F_2(p,q,t) = \frac{Jq(p,q,t)}{\rho(p,q,t)} \quad (2.9)$$

Integration of this system gives $p(t)$ and $q(t)$ which constitute quantum trajectory in phase space. Unlike classical equations of motion, system (2.9) has right side dependent on time. Moreover, each trajectory depends not only on the potential and initial condition, but also on the wave function. Having the latter for any moment of time one can derive $\rho$ and $Jq, Jp$ from (2.6)-(2.7).

Numerical integration of (2.9) can be performed if $F_1$ and $F_2$ are continuous function of $p, q$ because these variables change in time. This fact restricts application of numerical methods which use phase space discretization of wave function on a fixed grid.

Thus, two principal problems may arise in procedure of quantum trajectory deriving: non-stationary wave function obtaining and numerical integration of system (2.9).

## III. APPLICATION FOR 1D-POTENTIAL

Neither stationary nor non-stationary Schroedinger equation does have an analytical solution. We can use the standard approach which consists in expanding of initial state in eigenstates

$$\Phi(p,q,t) = \sum a_n \Psi_n(p,q) \exp(-iE_n t) \quad (3.1)$$

where

$$a_n = \int_{-\infty}^{\infty}\int_{-\infty}^{\infty} \Psi_n^*(p,q)\Phi(p,q,0)dpdq \quad (3.2)$$

Usually it takes a long computation time for deriving of $a_n$. Nevertheless, one can only rely on a superposition of two eigenstates, to explore some aspects of trajectory behavior. It is especially convenient for double well potential because superposition of even and odd neighbor states gives one which is located at one of the wells.

### A. Superposition of two lowest states

The quantum state

$$\Phi(p,q,t) = \frac{1}{\sqrt{2}}(\Psi_0(p,q)\exp(-iE_0 t) + \Psi_1(p,q)\exp(-iE_1 t)) \quad (3.1)$$

evolves periodically in time with tunneling period

$$\tau = \frac{2\pi}{E_1 - E_0}$$

For this time it passes through the barrier and goes back. After half-period of tunneling it completely concentrates in another well with antisymmetric shape.

At potential parameters $\alpha=0.1$ and $E_0=3$ there are four eigenstates whose energy is lower than barrier height. Linear combination of two nearest states at different time points is shown in Fig.3.1.

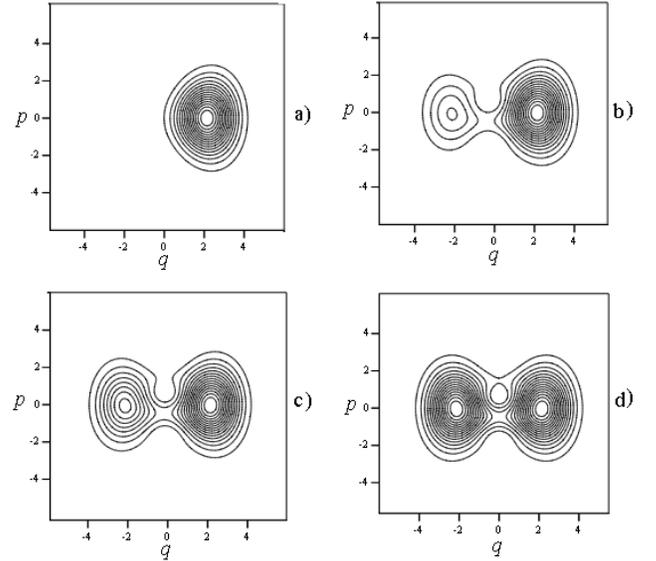

**Figure 3.1**. Probability density at different time points: a) t=0; b) t=0.1$\tau$; c) t=0.2$\tau$; d) t=0.25$\tau$.

Maximum density is at $p=0$, $q=2.27$ but energy minimum is at $p=0$, $q=2.34$. The period of tunneling is $\tau$ =957.76. Integration of (2.9) by second order difference scheme with initial condition $p(0)=0$, $q(0)=3$ gives the quantum trajectory shown in Fig.3.2-3.3. For time 6.95 a.u. the trajectory makes a complete revolution. Such behavior reminds a little classical one for particles whose total energy is less than barrier height (Fig.3.2a).

At the same time it has something unusual. Firstly, its classical energy does not conserve. This fact was also found in [4] for motion in harmonic oscillator potential. Secondly, the trajectory does not go to initial point after first revolution.

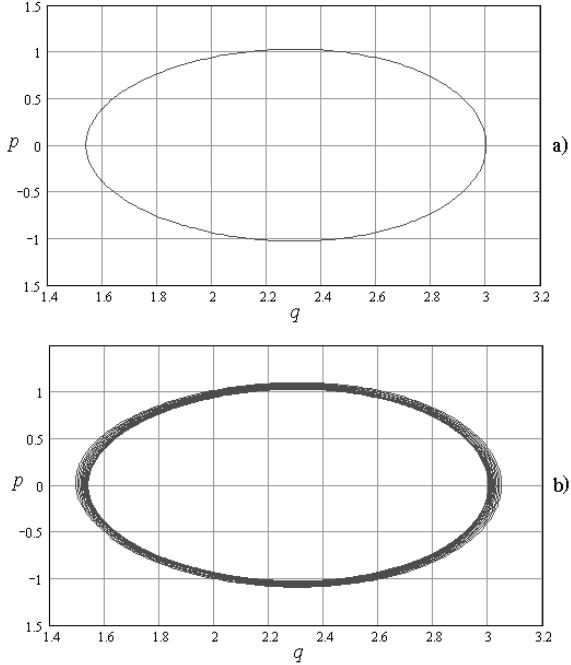

**Figure 3.2.** Quantum trajectories for time 6.95 a.u. (a) and 100 a.u. (b) at the same initial condition.

The coordinate at *p=0* after first circulation is 3.00022, after second one 3.00089, and so on. Although slightly, but the trajectory diverges with every revolution. This accumulation in small divergence results in trajectory spreading from the initial loop and finally in transition to another well at certain time (Fig.3.3). Thus tunneling occurs. The time before tunneling happens depends on initial condition. Usually if the initial condition is closer to the point of maximum probability density, the time of the trajectory staying in first well is longer. After tunneling, the trajectory continues to circulate in another well. But now this motion is directed inside. Such motion lasts until time is equal to the half period of tunneling. After reaching this time point the trajectory begins to move to outside. The process repeats but now the trajectory moves anti-symmetrically with respect to q-axis. This makes sense because the process is periodic as noted.

During passing through the barrier the trajectory becomes bent. It is interesting that this bend is not at q=0 but usually lies at 0<q<1 for all trajectories of this quantum state.

As seen from Fig.3.3., the classical separatrix does not influence quantum trajectory. For instance, starting at points with classical energy lower than barrier height the trajectory can cross the seperatrix few times without passing to another well. So no matter what the classical energy at initial point is. Tunneling is determined by initial quantum state and initial condition of trajectory.

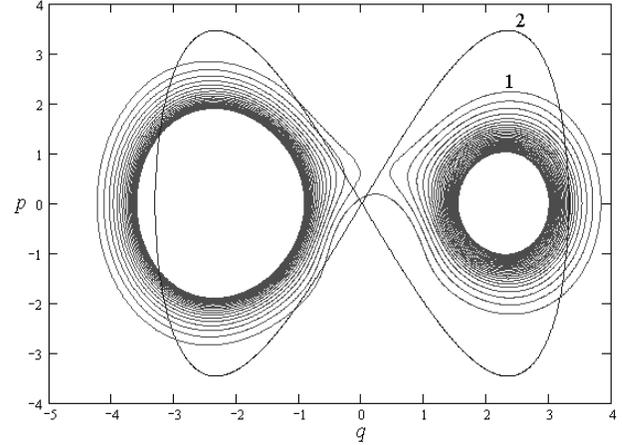

**Figure 3.3.** Quantum trajectory (1) for *0.5τ* time starting at $p_0=0$, $q_0=3$ and classical separatrix (2).

As a matter of fact, each trajectory tunnels in different ways. Therefore tunneling of single trajectory does not provide the important information about preferable tunneling path in phase space. To derive it one should analyze the swarm of trajectories. This approach is not effective due to the very long computational time. More appropriate parameter may be *-Jq(p,0,t)*. We call it as transferring rate, because its value is proportional to probability density transferring through the line of maximum potential barrier in phase space. Minus sign is set due to probability flow from positive to negative coordinates. In some cases, especially if the wave function is periodic in time, it is more rationally to use the integral characteristic

$$I(p) = \int_0^{0.5 Period} -Jq(p,0,t)dt$$

The graph in Fig.3.4 shows *I(p)*. As seen, the maximum almost coincides with *p=0*. Perhaps it is caused by the fact that maximum of probability is located at this line.

There are also two negative peaks. This is explained by the fact that some trajectories after crossing zero point *q*-line move clockwise and cross it again from the opposite side contributing to the negative part of transfer value (see Fig.3.3). Form of the graph in Fig.3.4 qualitatively is almost the same for all potential parameters $\alpha$ and $E_0$.

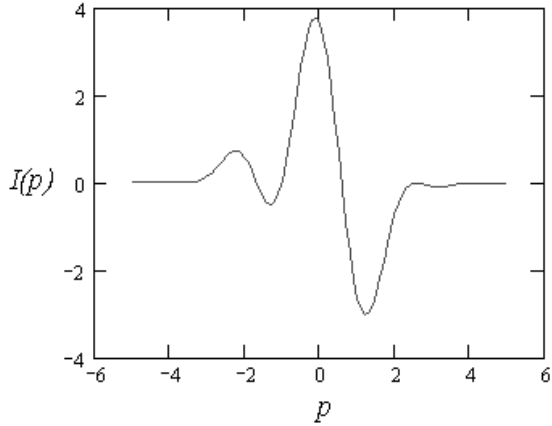

**Figure 3.4.** Transferring rate for different momenta.

### B. Gaussian wave packet

As noted above, the properties of quantum trajectory essentially depend not only on potential and initial condition as in classical dynamics but also on the quantum state. Thus there is a need of study of a problem for different states. In spite of relatively simple study of the state which is a superposition of two eigenstates, it has a shortcoming because the maximum density location and mean energy of quantum state are predicted parameters and cannot be freely varied.

The standard non-stationary method described above can be applied to any initial distribution. The most interesting is a study of the Gaussian wave packet

$$G(p,q) = \frac{1}{\sqrt{2\pi}} \exp\left(-\frac{(q-q_0)^2}{4\sigma^2} - \frac{\sigma^2(p-p_0)^2}{4}\right) \exp\left[\frac{i}{2}(p_0 q - p q_0)\right]$$

Its maximum can be placed anywhere in phase space, and its mean energy depends on it. If the barrier is too high, then Gaussian centered at $p_0=0$ and $q_0=q_{max}$, where $q_{max}$ is a point of maximum probability, is a superposition of two lowest eigenstates. Thus the task of the Gaussian study reduces again to the previous problem of superposition of two lowest states again.

As noted above, the mean energy of quantum state $E$ is determined by the initial condition of Gaussian. If its energy is higher or commensurable with the barrier height then the packet moves freely between both wells. Its initial shape does not preserve and even breaks to pieces [10]. In the harmonic oscillator potential such packet conserves its shape and the center moves according to classical laws of motion. It is interesting to figure out if the motion of the center in double well potential has a special feature as well.

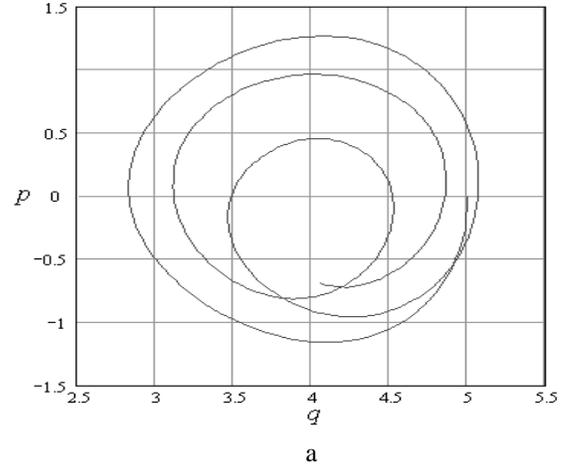

a

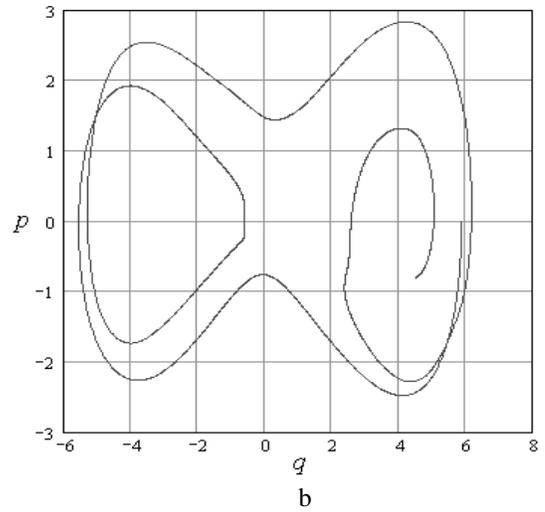

b

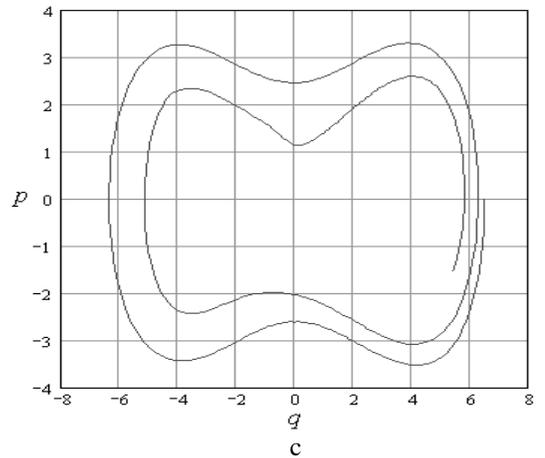

c

**Figure 3.5.** Quantum trajectories starting from maximum of wave packet which is located: a) inside the separatrix $p_0=0$, $q_0=5.0$, $E=1.42$; b) on the separatrix $p_0=0$, $q_0=5.89$, $E=4.13$; c) outside of the separatrix $p_0=0$, $q_0=6.5$, $E=6.651$.

The maximum packet was placed at the separatrix ($E_0=3$), inside it and outside. Then the quantum trajectories starting from maximum of packet were calculated. As seen from Fig.3.5, the quantum trajectories do not obey the classical law of motion anywhere. So, the probability maximum of Gaussian does not have any general special properties as in quadratic potential.

In classical mechanics there is a sharp difference between trajectories whose energies are lower and higher than the barrier height. In first case the motion is bound to one well. In the second case the trajectory moves around both wells.

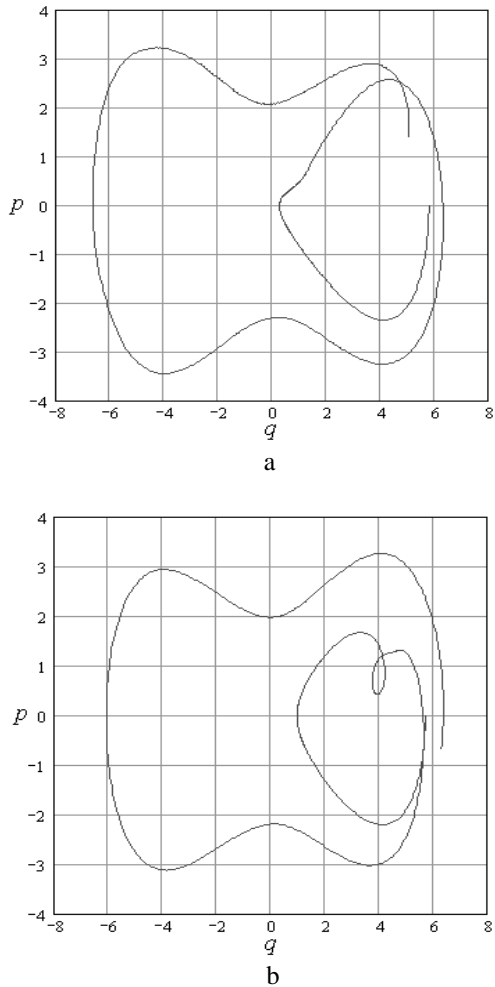

**Figure 3.6.** Quantum trajectories starting from maximum of wave packet which is located at: a) $p_0=0$, $q_0=5.79$; $E=3.66$; b) $p_0=0$, $q_0=5.70$; $E=3.37$.

Although the motion of quantum trajectory depends on mean energy of quantum state, there is no such sharp difference in behavior as in classical case. At relatively high mean energies of packet the trajectory passes to another well at once. Decreasing of energy leads to change in trajectory behavior. At energy 3.66 (see Fig.3.6) it makes one revolution in the first well. A number of revolutions in the first well continues to rise with mean energy decreasing. For instance, at $E=3.37$ it is 2. At very low energy, trajectory revolves in the first well for a long period of time. Unfortunately numerical integration of system (2.9) cannot be carried out for such a long time due to accumulated errors. Slow rising of the probability density in another well at earlier time indicates that some trajectories must pass through the barrier. Integration of (2.9) identifies them as trajectories with remote initial point from the maximum density point. Moreover, the more remote initial point of trajectory is, the less time for tunneling is needed (Fig.3.7).

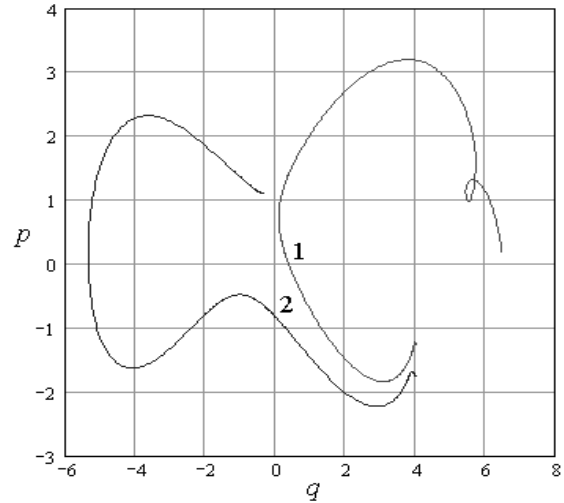

**Figure 3.7.** Trajectories starting from remote points: 1) $p(0)=-1.25$, $q(0)=4$; 2) $p(0)=-1.75$, $q(0)=4$. Initial Gaussian is centered at $p_0=1$, $q_0=4$; its energy $E=1.08$

Now we can conclude some general features for quantum dynamics in double well potential. If the starting point is further from maximum then it tunnels sooner, no matter what its classical energy is. Points with classical energy lower than barrier height may tunnel at once but ones with higher energy stay in first well crossing the separatrix many times from inside and outside. In the tunneling process distance between the maximum density and initial point plays an important role.

Another interesting point is to analyze tunneling rate. Fig.3.8. shows that the tunneling rate increases till certain time and then decreases. The peak moves from negative value of momentum to zero. Simultaneous vanishing of positive peak is accompanied by appearance of negative one at positive value of momentum. It indicates that trajectories return to first well.

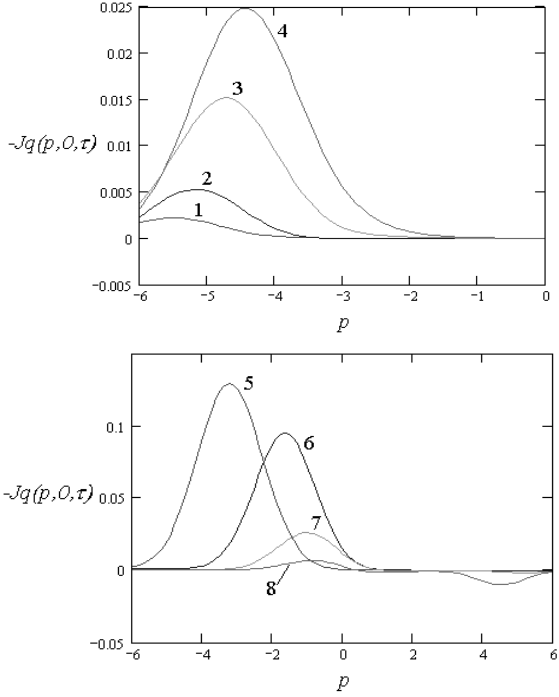

**Figure 3.8..** Tunneling rate for Gaussian at different time: 1) $\tau=1$; 2) $\tau=1.2$; 3) $\tau=1.4$; 4) $\tau=1.5$; 5) $\tau=2$; 6) $\tau=3$; 7) $\tau=4$; 8) $\tau=5$.

It is worth emphasizing that trajectory always moves clockwise. As a result, the Gaussian placed symmetrically to *p*-axis has different trajectories which start from symmetrical points although their energy is the same.

## IV. APPLICATION FOR 2D POTENTIAL

A system in two dimensions has been also considered. The potential has one degree of freedom as a double well and another as a harmonic oscillator. There is a linear or quadratic coupling between them (Fig.4.1). This type of potential has application in study of many chemical reactions. One of them is isomerization of malonaldehyde [11]. It has form for linear coupling

$$U(qx,qy) = \alpha qx^4 - \beta qx^2 + E_0 + \gamma qy^2 - \sigma qxqy \quad (4.1)$$

and for quadratic

$$U(qx,qy) = \alpha qx^4 - \beta qx^2 + E_0 + \gamma qy^2 - \sigma qx^2 qy \quad (4.2)$$

Coupling parameter $\sigma$ is responsible for shift of wells from the *qx*-axis.

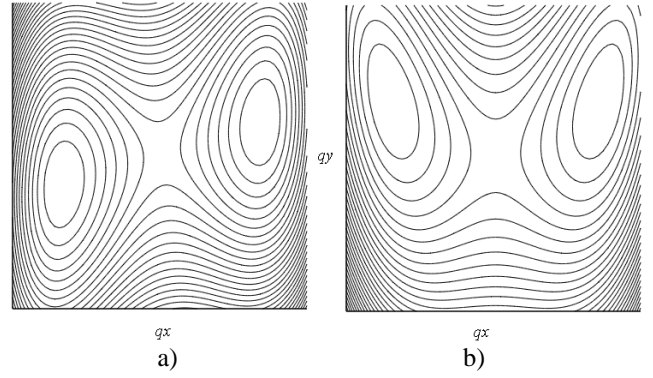

**Figure 4.1.** Contour plot of 2D potential: a) linear coupling; b) quadratic coupling.

As initial quantum state we used only superposition of the two lowest states. In this case, changing of coupling parameter leads not only to the shift of wells but also to change of eigenstates which constitute the initial quantum state. The more appropriate option would be to take the Gaussian as initial state because it does not depend on the coupling parameter. Unfortunately it cannot be used due to cumbersome calculation procedure in 2D.

In two dimensions the trajectory can be visualized using its projection on different surfaces. The probability can be transferred in *qx* and *qy* directions. Therefore the transferring rates such as *-Jqx(px,py,0,qy,t)* and *-Jqy(px,py,qx,0,t)* can be considered as function of *px, py, qy* or *px, py, qx* respectively. It is practically to reduce the number of variables making some of them fixed. For example, if it is necessary to have transferring rate in *qx-qy* surface, *px* and *py* should be fixed (usually zero) and *Ix=f(qy), Iy=f(qx)*.

$$Ix(qy) = \int_0^{0.5Period} -Jq(0,0,0,qy,t)dt$$

$$Iy(qx) = \int_0^{0.5Period} -Jq(0,0,qx,0,t)dt$$

In all following calculations the potential parameters in (4.1) and (4.2) are $\alpha=0.1$, $E_0=3$, $\gamma=0.5$, $\beta$ is set such a way that the potential minimum would be zero, $\sigma$ is varied from 0 to 0.3. Initial condition for the trajectory is $px_0=1$, $py_0=1$, $qx_0=3$, $qy_0=1$. As noted above, a long time is needed before tunneling happens. Therefore the initial time point for trajectory is $0.45\tau$.

### A. Zero coupling

At zero value of coupling parameter the two-dimensional problem reduces to the two one-dimensional tasks. Thus trajectory projections on *px-qx* and *py-qy* behave independently on each other. It is clear from the fact, that this potential is symmetric with respect to *qx*- and *qy*-axes. Therefore trajectory transition to another well does not influence form of *py-qy* projection, which is ellipse with center at origin. Transferring rate *Ix(px)* in *px-qx* surface has the same form as in one dimension. In *qy*-direction there is no tunneling.

### B. Linear coupling

If there is a linear coupling in potential, it is no more symmetrical relatively *qx* and qy axes though it is still symmetric with respect to origin. Value of sigma is responsible for the shifting of wells from *qx*-axis. Now the projections of potential minima on *qy*-axis do not coincide. Therefore motion in different wells should give two ellipse-like curves whose centers are near to potential minimum. It has to effect *py-qy* trajectory projection if transition to another well occurs. Fig.4.2 shows that there is a bend in *py-qy* curve when tunneling occurs. This bend is more distinct if the coupling is stronger. An increase of σ leads also to appearance of other bends in *py-qy* projection. They are typical for classical mechanics with this potential and usually correspond to sudden change of the trajectory direction in some projections. *px-qx* projection is almost uninfluenced by σ.

As seen from Fig.4.3, maximum of transferring rates in both directions is at zero point for any σ. It is due to the central symmetry of the potential. Thus preferable tunneling path lies along the straight line connecting wells. So it coincides with the line of minimum potential. As expected, transferring rate in *qy*-direction is unchanged (Fig.4.3a) but in *qx*-direction increases with increase of coupling parameter (Fig.4.3b) because both wells move away from each other in *qy*-direction.

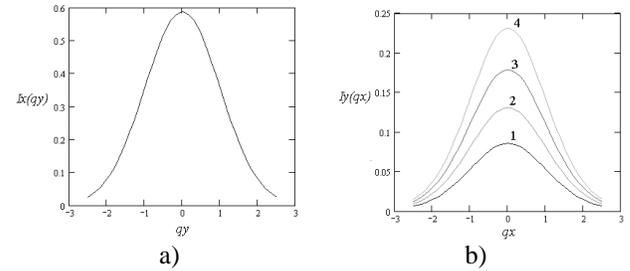

**Figure 4.3.** Transferring rate in: a) *qy*-direction; b) *qx*-direction. 1- σ=0; 2- σ=0.1; 3 - σ=0.2; 4 - σ=0.3

### C. Quadratic coupling

The problem is more complicated if the potential has quadratic coupling. Projection on *py-qy* surface has more bends. Now coupling raising influences on *px-qx* projection. It becomes more spreading in each well.

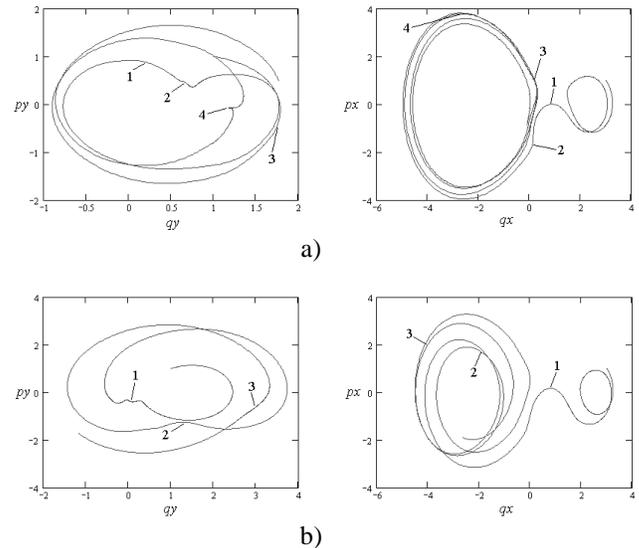

a)

b)

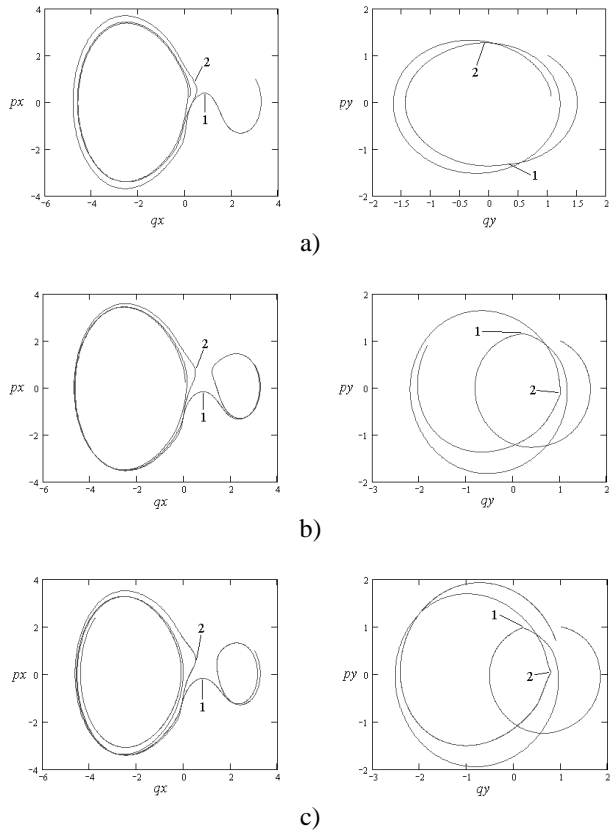

**Figure 4.2.** Projections *px-qx* and *py-qy* for the coupling parameter: a) σ=0.1; b) σ=0.2; c) σ=0.3. Numbers indicate points of same time.

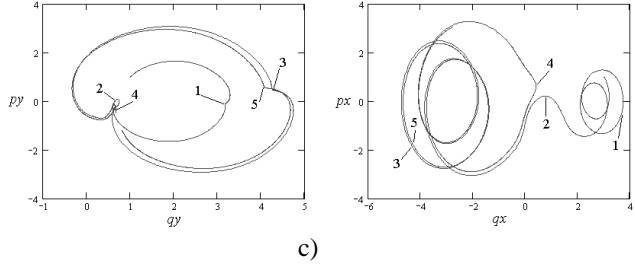

**Figure 4.5.** Projections *px-qx* and *py-qy* for coupling parameter: a) $\sigma=0.1$; b) $\sigma=0.2$; c) $\sigma=0.3$. Numbers indicate points of same time.

The preferable tunneling path is situated between line of minimum potential and straight line connecting the wells.

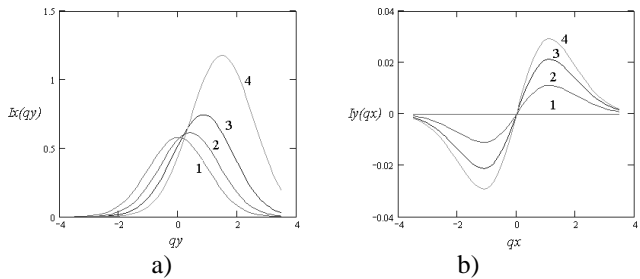

**Figure 4.3.** Transferring rate in: a) $qy$-direction; b) $qx$-direction. 1- $\sigma=0$; 2- $\sigma=0.1$; 3 - $\sigma=0.2$; 4 - $\sigma=0.3$

Increase of coupling causes raising and shifting of transferring rate maximum in *qy*-direction. In *qx*-direction it is antisymmetric with regard to *qx*=0.

## V. CONCLUSION

Unlike the classical mechanics the conservation of phase space volume in quantum mechanics fails [9]. This fact leads to more sophisticated structure of quantum trajectory. Quantum phase space structure is dependent not only on Hamiltonian but also on the initial quantum state. Therefore the classical energy and probability density along the trajectory do not conserve. As a result, the trajectory may cross the classical separatrix few times, while staying in the initial well. The conception of potential barrier loses its sense in quantum world. Sure, the barrier height effects (penetration delaying) quantum trajectory but it is no more the wall which cannot be penetrated.

The results obtained in this work concerning the description of quantum dynamics in double well potential have important information but they cannot be considered as absolute or complete because only two types of initial states were used. Other different initial states may give some new details.

## ACKNOWLEDGMENT

The author thanks John H.Frederick for helpful discussions and comments.

___

[1] E.Wigner, Phys. Rev. **40**, 749 (1932)
[2] M.Hillery, R.F.O'Connell, M.O.Scully, and E.P.Wigner, Phys. Rep. **106**, 121 (1984)
[3] K.Husimi, Proc. Phys. Math. Soc. Jpn. **22**, 264 (1940)
[4] Go.Torres-Vega, and J.H.Frederick, J. Chem. Phys. **98**, 3103 (1993)
[5] K.B.Moller, T.G.Jorgensen, and Go.Torres-Vega, J. Chem. Phys. **106**, 7228 (1997).
[6] Go. Torres-Vega, A.Zuniga- Segundo, and J.D.Moralez-Guzman, Phys. Rev. A **53**, 3792 (1996)
[7] Go.Torres-Vega, and J.H.Frederick, J. Chem. Phys. **93**, 3862 (1990)
[8] Xu-Guang Hu, and Qian-Shu Li, J. Phys. A: Math. Gen. **32**, 139 (1999)
[9] R.T.Skodje, H.W.Rohrs, and J.VanBuskirk, Phys. Rev. A **40**, 2894 (1989)
[10] Go.Torres-Vega, K.B.Moller, and A.Zuniga-Segundo, Phys. Rev. A **57**, 771 (1998).
[11] N.Makri, and W.H.Miller, J. Chem. Phys. **91**, 4026 (1989)